\documentclass[]{article}

\usepackage{graphicx}
\usepackage{dcolumn}
\usepackage{bm}
\usepackage[utf8]{inputenc}
\usepackage[T1]{fontenc}
\usepackage{mathptmx}
\usepackage{amsfonts}
\usepackage{bbm}
\usepackage[bottom]{footmisc}
\usepackage[utf8]{inputenc}
\usepackage{amsmath}

\fontsize{11}{16}\selectfont

\begin{document}

\title{Some mathematical issues regarding a new approach towards quantum foundations}

\author{Inge S. Helland, Department of Mathematics. University of Oslo\\ P.O. Box 1053, N-0316 Oslo, Norway\\ ingeh@math.uio.no\\ ORCID: 0000-0002-7136-873X\\ July 26, 2025 }

\date{}

\maketitle

\begin{abstract}
In this article, the weakest possible theorem providing a foundation for the Hilbert space formalism of quantum theory is stated. The necessary postulates are formulated, and the mathematics is spelt out in detail. It is argued that, from this approach, a general epistemic interpretation of quantum mechanics is natural. Some applications to the Bell experiment and to decision theory are briefly discussed. The article represents the conclusion of a series of articles and books on quantum foundations.
\end{abstract}


\section{Introduction}

In a number of articles, the newest ones being Helland (2024a,b,c), this author has proposed a completely new foundation for quantum theory, a foundation based upon theoretical variables, which in a given context may be attached to an observer or to a group of communicating observers. These variables can be accessible or inaccessible. An example of a (maximal) accessible variable may be the positions $\bm{q} = (\bm{q}_1, ... ,\bm{q}_n)$ of $n$ independent particles; another example may be their momenta $\bm{p} = (\bm{p}_1, ... ,\bm{p}_n)$. A typical inaccessible variable may be the vector $(\bm{q},\bm{p})$.

From a mathematical point of view, the notion of a theoretical variable will be considered primitive. The only requirement is that if $\theta$ is a theoretical variable and $\xi$ is a function of $\theta$, then $\xi$ is a theoretical variable. Similarly, the notion of an accessible variable is primitive. Again, it is only required that if $\theta$ is accessible and $\xi$ is a function of $\theta$, then $\xi$ is accessible.

Given these primitive notions, the theory may be implemented in different ways. In this article, I will concentrate on the case where the variables are physical variables like positions, momenta or spin components. A physical variable is said to be accessible if it can in principle be measured with arbitrary accuracy.

Other implementations are possible. In Helland (2023e) the theoretical variables are decision variables. In Helland (2025) they are statistical parameters. In both cases, a fairly rich applied theory can be constructed from the mathematical theory considered here.

By assuming the existence of two different maximal accessible variables, in Niels Bohr's terminology two complementary variables, and making some additional assumptions, it is shown in the above articles that essentially the whole Hilbert space apparatus results. The purpose of the present article is to look closer at the additional assumptions. It turns out that these can be considerably weakened.

What we do have to assume is that the two variables, called $\theta$ and $\eta$ in the general theory, may be seen as functions of some basic inaccessible variable $\phi$, and that groups act on both $\phi$ and $\theta$.

In Helland (2024a) it was assumed that the two actual variables were related ($\eta(\phi)=\theta(k\phi)$ for some transformation $k$ on the space $\Omega_\phi$, the range of $\phi$). It is shown here that we only have to assume: 1) The spaces $\Omega_\theta$ and $\Omega_\eta$ have the same category; 2) There is a group $M$ acting on $\Omega_\phi$. In the example above, we can let $M$ be the multivariate Weil-Heisenberg group.

We also have to assume that there is a transitive group $G$ acting on $\Omega_\theta$, and that a left-invariant measure $\mu$ with respect to $G$ is given. In the above example, we can just let $G$ be the translation group.

In Helland (2024a) it was assumed that a multivariate representation $U(\cdot)$ of $G$ with certain properties existed. In the present paper, we show how such a simple representation can be constructed.  This assumption is simply not necessary.

From this, the weakest possible version of my main theorem is given as Theorem 1 below. The conclusion of the theorem is that every accessible variable has a symmetric operator in $\mathcal{H}=L^2 (\Omega_\psi, \nu )$ attached to it, where $\psi=(\theta,\eta)$, and $\nu$ is an invariant measure on $\Omega_\psi$ induced by the invariant measure $\mu$ on $\Omega_\theta$. (Strictly speaking, for this conclusion we need technical assumptions such that the spectral theorem is valid for symmetric operators, not only for self-adjoint operators.) This is the starting point for much of the Hilbert space apparatus. Apart from the above symmetry assumption, the essential assumption is only the existence of two complementary theoretical variables in the given context.

Note that there are no microscopic assumptions here. Thus, this derivation also gives a foundation for what Khrennikov (2010, 2023) calls quantum-like models. These models have links to several scientific disciplines. The link to quantum decision theory will be discussed elsewhere. Links to relativity theory and quantum field theory are discussed in Helland (2023c) and in Helland and Parthasarathy (2024).

I will concentrate on the derivation of the Hilbert space apparatus and related derivations in this article. Assumptions that lead to the Born rule for probabilities are discussed, and the derivation is proved, in Helland (2024b). The assumptions are also discussed in Section 3 below. A derivation of the Schr\"{o}dinger equation from a few postulates has been given for instance by Klein (2010).

Finally, it is a basic setting in that the theoretical variables are attached to an observer or to a group of communicating observers. A natural assumption in addition is that they are connected to the mind(s) of this/these observer(s). This leads to a general epistemic interpretation of quantum theory, an interpretation that contains QBism (see for instance Fuchs et al., 2013) as a special case. Quantum theory is seen as a theory of our knowledge about the real world, not directly about the real world. This is discussed in Section 4.

The plan of this article is as follows: in Section 2 the theory is outlined in its weakest possible version. In Section 5 some consequences, consequences for an understanding of the Bell theorem and for a new theory of decisions, are briefly discussed. In Section 6 some final remarks are given.

\section{The main theorems}

I repeat that my main notion is that of theoretical variables, which can be almost anything. The theoretical variables can be accessible or inaccessible. 

For physical modelling I assume a fixed context, and that an observer or a group of communicating observers in this context has/have a set of theoretical variables associated with him/them. In the case of a group of observers, their communication should be related to these variables. Then my first postulate is as follows:
\bigskip

\textbf{Postulate 1}

\textit{There is an inaccessible variable $\phi$ such that all accessible variables can be seen as functions of $\phi$. There is a group $M$ acting on $\Omega_\phi$.}
\bigskip

In simple physical examples, such a $\phi$ can easily be constructed. As a general statement covering all possible situations, Postulate 1 can also be given a religious interpretation; see Helland (2022d, 2023d). This will be further discussed elsewhere.

One possible option is to replace Postulate 1 with some assumptions in category theory; see the arXiv version of Helland (2024a). Category theory in the foundation of quantum mechanics has also been considered by others, for instance, Coecke and Papette (2009) or Döring and Isham (2008). This option will not be considered further in the present article.

My main theorems will refer to a situation where we have two different maximal accessible theoretical variables, which I, following Niels Bohr, will call two complementary variables. I will show that the whole Hilbert space apparatus follows under weak conditions from the assumption that we have two such maximal accessible variables. The term `maximal' means roughly that the variable cannot be extended and still be accessible. To be precise, I need to define a partial ordering among the variables.
\[\]
\[\]

\textbf{Definition 1}

\textit{Say that  $\theta\le\lambda$ if $\theta = f(\lambda)$ for some function $f$.}
\bigskip

This is a partial ordering among all theoretical variables and also a partial ordering among the accessible ones. Note that $\phi$ from Postulate 1 is an upper bound in the accessible case. (Again, one must make precise what is meant by a function. For instance, one can let $\Omega_\phi$ and $\Omega_\theta$ be topological spaces and concentrate on functions that are Borel-measurable.) We will say that $\theta$ is a maximal accessible variable if it is maximal with respect to this partial ordering. By Zorn's Lemma, which is equivalent to the Axiom of Choice, maximal variables always exist. For those who do not believe in Zorn's Lemma, we add an additional postulate.
\bigskip

\textbf{Postulate 2}

\textit{For every accessible variable $\xi$ there is a maximal accessible variable $\theta$ such that $\xi\le\theta$.}
\bigskip 

In order to achieve a meaningful theory, we also need some symmetry assumptions. One such is given by the existence of a group $M$ acting upon $\phi$ (Postulate 1). Another assumption is given by the existence of a group $G$ acting upon $\theta$. In concrete examples, these groups can be easily constructed.
\bigskip

\textbf{Postulate 3}

\textit{To a given accessible theoretical variable $\theta$ there is a group $G$ acting upon $\theta$, and there is a left-invariant measure $\mu$ with respect to $G$ on $\Omega_\theta$, the range space of $\theta$. The group $G$ is transitive and has a trivial isotropy group.}
\bigskip

 Conditions for the existence of an invariant measure are discussed in Helland (2021). Note that if an invariant measure is supposed to act on every single theoretical variable, there is consequently an invariant measure on every set of theoretical variables.
 
 Finally, in this article, I will assume for two complementary variables $\theta$ and $\eta$:
 \bigskip
 
 \textbf{Postulate 4}
 
 \textit{The range space $\Omega_\theta$ is either finite and has the same number of values as $\Omega_\eta$, or, more generally, $\Omega_\theta$ and $\Omega_\eta$ have the same category.}
 \bigskip
 
 (In terms of category theory, this means that $\Omega_\theta$ and $\Omega_\eta$ are objects, and that there is a morphism from $\Omega_\theta$ to $\Omega_\eta$ and another morphism from $\Omega_\eta$ to $\Omega_\theta$. More intuitively, it means that there is a bijective function connecting $\Omega_\theta$ and $\Omega_\eta$.)
 
 This is all we need for our first results: a Proposition and a basic Theorem.
 \[\]
 \[\]
 
 \textbf{Proposition 1}
 
 \textit{Assume that the basic inaccessible variable $\phi$ satisfies Postulate 1, and that two given accessible variables $\theta$ and $\eta$ satisfy Postulate 4. Then $\theta$ and $\eta$ are either in one-to-one correspondence, or the following holds: there exists an accessible variable $\xi$ which is a bijective function of $\eta$, a transformation $k$ in $\Omega_\phi$, and a function $f$ acting on $\Omega_\phi$ such that $\theta = f(\phi)$ and $\xi = f(k\phi)$.}
 \bigskip
 
 In many applications, it turns out that this will hold with $\xi$ equal to $\eta$. In that case we say that $\theta$ and $\eta$ are \emph{related}: $\theta=f(\phi)$ and $\eta=f(k\phi)$ for some $k$.
 \bigskip
 
 \underline{Proof}
 
 The finite-dimensional case was treated in Section 7.2 of Helland (2024a), so I will here look at the more general case. Choose a function $f$ such that $\theta=f(\phi)$, and fix $\phi=\phi_1$. Let $\phi_2$ be any point in $\Omega_\phi$ such that $\eta(\phi_1)=f(\phi_2)$. Such a $\phi_2$ must exist, since $\{\eta(\phi)\}$ has the same category as  $\{\theta(\phi)\}=\{f(\phi)\}$.
 
 The group $M$ acting upon $\Omega_\phi$ need not be transitive. The points $\phi_1$ and $\phi_2$ either lie on the same orbit of $M$ or on different orbits. In the first case, there exists a $k\in M$ such that $\phi_2 =k\phi_1$, so that $\eta(\phi_1)=f(k\phi_1)$. In the second case, let $a$ be a function of $\phi$ which characterises the orbits. Then there exists a $\phi_3$ on the orbit containing $\phi_1$ such that $\phi_2 =a(\phi_3)$, and by definition $\phi_3 = k\phi_1$ for some $k\in M$. In this case we get $\eta(\phi_1) =f(\phi_2) = f(a(\phi_3))=f(a(k\phi_1))$. Define $\xi (\phi_1)= f(k\phi_1)$. Then $\eta(\phi_1) =f(a(f^{-1}(\xi(\phi_1))))$, which means that $\eta$ is a bijective function of $\xi$. The inverse $f^{-1}$ is well defined, since, by $\xi (\phi_1)= f(k\phi_1)$, and letting $\phi_1$ vary, the range of $\xi$ has the same category as the range of $\eta$.
 
 Since this holds for every $\phi_1\in \Omega_\phi$, the Proposition is proved.
 
$\Box$
\bigskip

The Theorem, which is a refinement of the basic Theorem 4 in Helland (2024a), runs as follows
\smallskip

\textbf{Theorem 1}

\textit{Assume Postulate 1 and Postulate 2. Let $\theta$ and $\eta$ be two maximal accessible variables satisfying Postulate 4 that are not in one-to-one correspondence, and assume that they are real-valued or real vectors. Let $\theta$ satisfy Postulate 3. Then there exists a Hilbert space $\mathcal{H}$, and to every accessible variable $\zeta$ there exists a unique symmetric operator $A^\zeta$ in $\mathcal{H}$.}
\bigskip

\underline{Proof}

By Proposition 1 there exists a maximal variable $\xi$ such that $\xi$ and $\theta$ are related. Then it follows from Theorem 4 of Helland (2024a) (see also Theorem 1 of Helland, 2022a) that there exists a Hilbert space $\mathcal{H}$ such that every accessible variable is associated with a unique symmetric operator in $\mathcal{H}$ if the following condition holds:
\bigskip

\textit{There exists a unitary multi-dimensional representation $U(\cdot)$ of $G$ such that for some $|\theta_0\rangle$ the coherent states $U(g)|\theta_0\rangle$ are in one-to-one correspondence with the values of $g\in G$ and hence with the values of $\theta$.}
\bigskip

I will now define a simple representation $U(\cdot)$ on the Hilbert space $L^2 (\Omega_\theta , \mu)$ satisfying this condition.
\bigskip

\textbf{Proposition 2}

\textit{For $f\in L^2 (\Omega_\theta , \mu)$ and $g\in G$, define $U(g)f(\theta)=f(g^{-1}\theta)=h(\theta)$. Then the above condition holds.}
\bigskip

\underline{Proof of Proposition 2}
\bigskip

\textbf{Lemma 1} 

\textit{$h\in L^2 (\Omega_\theta , \mu)$.}
\bigskip

\underline{Proof}

By the invariance, $\int_{\Omega_\theta} |f(g^{-1}\theta)|^2 d\mu =\int_{\Omega_\theta} |f(\theta)|^2 d\mu < \infty$.
$\Box$
\bigskip

\textbf{Lemma 2}

\textit{The mapping $g\rightarrow U(g)$ is a homomorphism.}
\bigskip

\underline{Proof}
\[U(g_1)U(g_2)f(\theta)=U(g_1)f(g_2^{-1}\theta)=f(g_2^{-1}g_1^{-1}\theta)=f((g_1g_2)^{-1}\theta).\]
\[U(g^{-1})f(\theta)=f(g\theta)=U(g)^{-1}f(\theta).\]
$\Box$
\bigskip

\textbf{Lemma 3}

\textit{$U(g)$ is unitary for every $g\in G$.}
\bigskip

\underline{Proof}

\[\int_{\Omega_\theta} (f_1(\theta)^*U(g)^\dagger) f_2(\theta)d\mu= \int_{\Omega_\theta} f_1(\theta)^*(U(g)^{-1} f_2(\theta))d\mu.\]
$\Box$
\bigskip

\textbf{Lemma 4}

\textit{Choose $f_0 \in L^2 (\Omega_\theta , \mu)$ such that $f_0 (\cdot)$ is a bijective function of $\theta$. Then there is a one-to-one correspondence between $g\in G$ and the coherent functions $f_g (\theta)=U(g)f_0(\theta)=f_0 (g^{-1}\theta)$.}
\bigskip

\underline{Proof}

$f_0(g_1^{-1}\theta)= f_0(g_2^{-1}\theta)$ implies $f_0(g^{-1}\theta)=f_0(\theta)$ for $g=g_1^{-1}g_2$. Since $f_0$ is bijective, it follows that $g=e$. $\Box$
\bigskip

Theorem 1 now follows from Theorem 4 of Helland (2024a) and Proposition 2. $\Box$
\bigskip

This result, together with the other results of Helland (2024a,b,c), now gives a very simple alternative foundation of quantum theory. Note that this is a purely mathematical theory, and it can be interpreted in different directions. In an ordinary physical setting, it is natural to interpret the accessible theoretical variables as ordinary physical variables, but also as connected to the mind of an observer or to the joint minds of a communicating group of observers. This gives a general epistemic interpretation of quantum theory, an interpretation which has QBism as a special case. Quantum theory is then a theory of an observer's or a group of observers' knowledge about the real world, not a theory directly about the real world.

In this article, I will not go into detail with the results of the articles mentioned above, but one mathematical result deserves to be mentioned.
\[\]

\textbf{Theorem 2}

\textit{If $\theta$ and $\eta$ are related through a transformation $k$ of $\Omega_\phi$, then there exists a unitary operator $S(k)$ such that $A^\eta=S(k)^\dagger A^\theta S(k)$.}
\bigskip

\underline{Proof}

See Theorem 5 of Helland (2024a). $\Box$
\bigskip

I should also mention again the consequences for accessible theoretical variables that take a discrete set of values.

- Every accessible variable has a symmetric operator associated with it.

- The set of eigenvalues of an operator is equal to the possible values of the variable.

- An accessible variable is maximal if and only if all eigenvalues are simple.

- The eigenvectors can, in the maximal case, be interpreted in terms of a question together with an answer. Specifically, it means in a context with several variables, a chosen variable $\theta$ may be associated with a question `What is the value of $\theta$?' or `What will $\theta$ be if we measure it?', and a specific eigenvector of $A^\theta$, corresponding to the eigenvalue $u$, may be identified with the answer `$\theta=u$'.

- In the general case, eigenspaces have the same interpretation.

- The operators of related variables are connected by a unitary similarity transform.
\bigskip

In Theorem 1 it was concluded that the relevant operators were symmetric. This is a simple property: $\langle u|Av\rangle = \langle Au|v\rangle$ for all $|u\rangle , |v\rangle$ in the domain of $A$. To use the spectral theorem in general, we need operators corresponding to the two maximal accessible variables $\theta$ and $\eta$ to be \emph{self-adjoint}. Then the spectral theorem is valid (Hall, 2013), and it can be used to define operators corresponding to other accessible variables.

Look at the cases $A=A^\theta$ and $A=A^\eta$. Then we can recall the formulae (9) and (10) in Helland (2024a):
\begin{equation}
A^\theta = \int f_\theta(n)|v_n\rangle\langle v_n| \nu(dn),
\label{Atheta}
\end{equation}
\begin{equation}
A^\eta = \int f_\eta(n)|v_n\rangle\langle v_n| \nu(dn),
\label{Aeta}
\end{equation}
where $n\in N$, a group acting on $\psi = (\theta,\eta)$, $\nu$ is a left-invariant measure on $\Omega_\psi$, $f_\theta(n)=\theta$, $f_\eta(n)=\eta$, $|v_n\rangle = W(n)|v_0\rangle$ with $W(\cdot)$ being an irreducible representation of $N$, and $\int |v_n\rangle\langle v_n| \nu(dn)=I$.

The domain $D$ of $A^\theta$ is the set $|u\rangle\in \mathcal{H}$ where the integral $A^\theta |u\rangle$ converges. The domain $D^\dagger$ of its adjoint is the set $|u\rangle\in\mathcal{H}=L^2(\Omega_\psi ,\nu)$ such that the functional $\langle u|A^\theta \cdot\rangle$ is bounded, and for $|u\rangle\in D^\dagger$ the adjoint is defined by the requirement that $A^{\theta\dagger}|u\rangle$ is the unique vector $|w\rangle$ such that $\langle w|v\rangle =\langle u|A^\theta v\rangle$ for all $|v\rangle\in D$.

On investigating the self-adjointness of the symmetric operator $A^\theta$, the main work lies in proving that $D^\dagger = D$. Let $|u\rangle\in D^\dagger$. Then $A^\theta|u\rangle$ must be defined, so we always have that $D^\dagger\subseteq D$. Let  then $|u\rangle\in D$. Then by the symmetry also the integral $\langle w|=\langle u|A^\theta$ converges. The problem is to find conditions such that $|v\rangle\rightarrow \langle w |v\rangle$ defines a bounded functional. This means that there exists a constant $C$ such that $|\langle w |v\rangle |=|\langle v |A^\theta u\rangle |\le C\||v\rangle\|$ for all $|v\rangle\in D$. A sufficient condition for this, given (\ref{Atheta}), is 
\[\]

\textbf{Postulate 5}

\textit{The integral}
$ \int |f_\theta(n)|\langle u|v_n\rangle\langle v_n|u\rangle \nu(dn)$
\textit{converges for every $|u\rangle\in D$.}
\bigskip

Note that for $|u\rangle\in D$, the corresponding integral without absolute values converges.
\[\]

\textbf{Proposition 3}

\textit{If Postulate 5 holds, then $A^\theta$ is self-adjoint. A corresponding condition holds for $A^\eta$.}
\bigskip

\underline{Proof}

By the Cauchy-Schwarz inequality
\[|\langle v |A^\theta u\rangle |^2 \le  \int |f_\theta(n)|\langle u|v_n\rangle\langle v_n|u\rangle  \nu(dn) \int |f_\theta(n)|\langle v|v_n\rangle\langle v_n|v\rangle  \nu(dn),\] 
and the last two integrals are finite when $|u\rangle\in D$ and $|v\rangle\in D$. Without loss of generality assume $\||u\rangle\| = \||v\rangle\| = 1$. $\Box$
\[\]

The spectral theorem implies:
\bigskip

\textbf{Theorem 3}

\textit{ For maximal variables $\theta$ and $\eta$ that are not bijective functions of each other, the corresponding operators $A^\theta$ and $A^\eta$ do not commute.}
\[\]

\underline{Proof}

I will prove this for the case of discrete-valued variables. The spectral theorem then gives
\[ A^\theta = \sum_j \lambda_j \bm{v}_j \bm{v}_j^\dagger , \]
\[ A^\eta = \sum_i \mu_i \bm{u}_i \bm{u}_i ^\dagger. \]
Since $\theta$ is maximal, all the eigenvalues are different, so $A^\theta$ uniquely determines the set of eigenvectors $\{\bm{v}_j\}$ up to phase factors. Similarly, $\{u_i\}$ is uniquely determined by $A^\eta$. The two sets of eigenvectors satisfy $\sum_j \bm{v}_j \bm{v}_j^\dagger =I$ and $\sum_i\bm{u}_i \bm{u}_i ^\dagger =I$. These two sets of eigenvectors cannot be identical, for in this case $\theta$, taking the values $\lambda_j$, and $\eta$, taking the values $\mu_i$ would be bijective functions of each other. But when at least one $\bm{v}_j$ differs from the set of vectors $\{\bm{u}_i\}$, it follows from the formulae above that $A^\theta$ and $A^\eta$ do not commute. $\Box$

\section{Quantum probabilities}

For the derivation of Born's formula, two more postulates are needed.
\bigskip

\textbf{Postulate 6}

\textit{The generalised likelihood principle from statistics holds.}
\bigskip

The basis for nearly all statistical inference is a statistical model, a model for the data $z$ in terms of the total parameter $\theta$. This is expressed by a probability function $p(z|\theta)$, in the discrete case a point probability, and in the continuous case a probability density. The likelihood of the data is defined as $L(\theta |z)=p(z|\theta )$, the probability function seen as a function of the parameter. 

The following principle can be taken as the basis for very much statisitical inference, and in my theory it is also used as a motivation behind Born's formula.
\[\]

\textbf{The generalised likelihood principle}

\textit{Consider two experiments in the same context $\tau$, and assume that $\theta$ is the same full parameter in both experiments. Suppose that two observations $z_1$ and $z_2$ have proportional likelihoods in the two experiments, where the proportionality constant $c$ is independent of $\theta$. Then, these two observations produce the same experimental evidence on $\theta$ in this context.}
\bigskip

The term ``experimental evidence'' is here left undefined and can be specified in any desirable direction. In some statistical textbooks and articles, the likelihood principle is formulated without specifying a context. As discussed in Helland (2021, 2024b), specifying $\tau$ implies that the principle becomes much less controversial. One may also assume that the two contexts are similar, i.e., one-to-one functions of each other.

The second necessary postulate is
\bigskip

\textbf{Postulate 7}

\textit{Consider a physical context $\tau$ observed by an observer $B$ whose decisions are influenced by a superior actor $D$. Assume that $D$'s probabilities are taken as experimental evidence and that $D$ is seen by $B$ to be rational in agreement with the Dutch Book Principle.}
\bigskip

\textbf{The Dutch Book Principle}

\textit{No choice of payoffs in a series of bets shall lead to a sure loss for the bettor.}
\bigskip

From these two postulates, assuming perfect experiments in the sense that no experimental noise is assumed, and using a version of Gleason's theorem given by Busch (2003), the following version of Born's formula is proved.
\bigskip

 Consider two experiments with maximal accessible discrete parameters $\theta^a$ and $\theta^b$, having the same number of values. By a corollary to Theorem 1 (Helland, 2024a), there are operators $A^a$ and $A^b$ in a Hilbert space $\mathcal{H}$ associated with these two parameters; the eigenvalues of the operators are the possible values of the parameters, and the corresponding normed eigenvectors are simple. Let $|a;k\rangle$ correspond to the event $\theta^a = u_k$ and $|b;j\rangle$ correspond to the event $\theta^b =v_j$.
\bigskip

\textbf{Theorem 4}

\textit{The conditional probabilities are given by}

\[ P(\theta^b = v_j |\theta^a = u_k ) = |\langle a;k |b:j\rangle |^2 . \]

Thus, the probabilities are given by the squared norms of the probability amplitudes $c = \langle a;k |b:j\rangle $.
\bigskip

From this, more general variants of the Born formula can be proved, for instance the following, valid both for discrete and continuous variables, and taking as a point of departure a density operator $\rho^a$ computed from an assumed probability distribution over $\theta^a$:
\bigskip

\textbf{Theorem 5}

\[ E(\theta^b | \rho^a ) = \mathrm{trace} (\rho^a A^b ). \]
\bigskip

A referee has asked the following question: Why did we get precisely the complex Hilbert space formalism, and not the generalised probability theory, in the spirit of Khrennikov et al. (2025)? The generalised probability theory is described in detail in Barrett (2007) and M\"{u}ller (2021), but we do not need the definitions here.

The answer to why quantum probabilities in my theory are given by the Born formula, and not by any more generalised formula, is given by the proofs of the results above, and in particular by a consequence of the generalised likelihood principle. In the discrete case, one can define the likelihood effects

\begin{equation}
F^b (\bm{v}; z^b, \tau ) = \sum_j p(z^b |\tau , \theta^b = v_j )|b;j\rangle\langle b;j |, 
\label{effect}
\end{equation}
where again $|b;j\rangle$ corresponds to the event $\theta^b =v_j $. These likelihood effects are crucial in the proof of Theorem 4. They are closely related to the positive operator-valued measures (POVM) of traditional quantum theory, where for discrete data the measure of the event $z^b =z$ is given by $\sum_j p(z^b=z |\tau , \theta^b = v_j )|b;j\rangle\langle b;j |$.

Thus, my approach towards probabilities leads to traditional quantum probabilities.

\bigskip

It should also be mentioned that, in addition to the postulates of this article, a final postulate is needed to compute probabilities of independent events. A version of such a postulate is
\bigskip

\textbf{Postulate 8}

\textit{If the probability of an event $E_1$ is computed by a probability amplitude $c_1$ from the Born rule in the Hilbert space $\mathcal{H}_1$, the probability of an event $E_2$ is computed by a probability amplitude $c_2$ from the Born rule in the Hilbert space $\mathcal{H}_2$, and these two events are independent, then the probability of the event $E_1 \cap E_2$ can be computed from the probability amplitude $c_1c_2$, associated with the Hilbert space $\mathcal{H}_1\otimes\mathcal{H}_2$.}
\bigskip

This postulate can be motivated by its relation to classical probability theory: If $P(E_1)=|c_1|^2$ and $P(E_2)=|c_2|^2$, then

\[ P(E_1\cap E_2)=P(E_1)P(E_2)=|c_1|^2 |c_2|^2 = |c_1c_2|^2.\]

\section{The epistemic interpretation and Ozawa's intersubjectivity theorem}

The interpretation of quantum theory is still a theme of intensive debate and some confusion. In a recent poll among a group of  physicists (Jodlicska et al. 2025), 21\% supported an epistemic interpretation, while 24\% supported an ontic interpretation. 15\% supported a mix of the epistemic and the ontic interpretation, while 41\% said that they supported an ensemble interpretation. In Helland (2019, 2021), I gave several arguments for the epistemic viewpoint. These arguments were largely of non-mathematical nature, and so, the main discussion there is outside the scope of the present article. But one special issue in that connection contains some important mathematics.

In Ozawa (2019), the author proved the following: Assume two spacelike measurements $M_1$ and $M_2$ at times $\tau_1$ and $\tau_2$ of the same observable, associated with the operator $A$. Let $U(t)$ be the time evolution operator of the total system, i.e., physical system plus environment. Define $A(0) = A\otimes I$, $M_1(t) =U(t)^\dagger (I\otimes M_1 )U(t)$ and $M_2 (t) = U(t)^\dagger (I\otimes M_2)U(t)$. Assume that for all $x$
\begin{equation}
P(M_1 (\tau_1)=x) = P(M_2 (\tau_2)=x) = P(A(0)=x).
\label{reproduce}
\end{equation}
Then
\begin{equation}
P(M_1(\tau_1) =x, M_2(\tau_2)=y) = 0
\label{ intersubj}
\end{equation}
for $x\ne y$.

In other words, under the assumptions of space separability and a weak reproducibility condition, the two observers are forced to get equal measurements.

Khrennikov (2024a) recently argued that the QBism's fundamental statement that `the measurement of an observable is personal' is in contravention of Ozawa's theorem. His starting point is the following citation from Fuchs and Schack (2014): `The fundamental primitive of QBism is the concept of experience. According to QBism, quantum measurement is a theory that any agent can use to evaluate her expectations for the content of their experience.'

Following Khrennikov, Ozawa's theorem implies that, for accurate local observations, the measurement's outcome is intersubjective which is a strong objection to QBism. There is nothing concerning personal experiences when different observers by necessity obtain the same measurement result.

This should be contrasted with my own interpretation, which can be seen as a consequence of the mathematical theory sketched in Sections 2 and 3 above. This mathematical theory can, in a physical context be interpreted in two ways: First, the mathematics, including the Born formula, can be seen as what is given by the observations of a single observer. It should be emphasised that the probabilities in Born's formula can be seen as probabilities calculated by the hypothetical superior actor $D$. Next, the mathematics can be interpreted as the joint observations as seen by a group of communicating observers.

Note that, in Ozawa's theorem, the two observers are spacelike separated and do not communicate about their observations. However, spacelike separated observers may later meet and share their experiences. Then it is perfectly allowable that they find that their measurement values are equal.

In conclusion, there seems to be no contravention between my general epistemic theory and Ozawa's intersubjectivity theorem.

It is interesting that Khrennikov (2024b) argues that Rovelli's well-known relational quantum mechanics can be specified in a way where there is no contradiction to Ozawa's theorem. Relational quantum mechanics states that every statement about a system should be related to another system. This `other system' may well be an observer or a communicating group of observers, which provides a link to my interpretation.

It is important that my basis is a purely mathematical theory, and this theory can be interpreted in different directions. I stated above that it in a natural way leads to a general epistemic interpretation, connected to the knowledge of an observer or to the joint knowledge of a communicating group of observers. However, in some contexts, this group of communicating observers can be imagined to consist of all relevant persons in the world. In these contexts, we might as well consider an ontological interpretation of the actual aspects of quantum mechanics.

\section{A mathematical consequence and an interpretation}

This section has been included to illustrate an important aspect of my theory: From purely mathematical postulates and theorems, and by just adding a natural interpretation, the following conclusions are derived: 1) general psychological statements having universal validity; 2) an explanation of a physical phenomenon which has been verified empirically, but which otherwise seems to be difficult to understand. The theme of this section has been treated in previous articles, and the mathematical proofs are deferred to these articles. However, the above aspect of the theory has not been clearly stressed before.

The property of being related ($\eta(\phi)=\theta(k\phi)$ for some $k$), is an important relation between two maximal accessible theoretical variables $\theta$ and $\eta$. By Theorem 2, if $\theta$ and $\eta$ are related, there is a unitary similarity transformation between the corresponding operators. This theorem has an inverse for finite-dimensional variables.
\bigskip

\textbf{Theorem 6}

\textit{Consider two maximal accessible finite-dimensional theoretical variables $\theta$ and $\eta$. If there is a transformation $k$ in $\Omega_\phi$ and a unitary transformation $W(k)$ such that $A^\eta = W(k)^\dagger A^\theta W(k)$, then $\theta$ and $\eta$ are related.}
\[\]
\[\]

\underline{Proof}

Since, by Postulate 1, $\theta$ and $\eta$ are functions of $\phi$, the transformation $k$ induces a transformation $s$ on $\psi = (\theta,\eta)$. The theorem then follows from the Lemma of Section 3 in Helland (2023b). $\Box$
\bigskip

In my basic theory, I have assumed that the theoretical variables are associated with an observer or with a communicating group of observers in a given context. Concentrate here on the first case, and call the observer $O$. Then we have
\bigskip

\textbf{Theorem 7}

\textit{Assume that two finite-dimensional related maximal accessible $\theta$ and $\eta$ are associated with $O$ in some fixed context. Then $O$ cannot in the same context be associated with another maximal accessible variable $\lambda$ which is related to $\theta$, but not related to $\eta$.}
\bigskip

\underline{Proof}

This is a consequence of Theorem 6. See the proof of Theorem 1 in Helland (2023b). $\Box$
\bigskip

The assumption that the variables are connected to the same context is crucial. In my interpretation of quantum theory, I connect the variables to the mind of an observer or to the joint minds of a communicating group of observers. Then we can consider maximal observations made in some fixed context, which also means some fixed time.
\bigskip

\textbf{Corollary 1}

\textit{Assume that $O$ has two related finite-dimensional maximal accessible variables $\theta$ and $\eta$ in his mind at some fixed time $t$. Then he cannot simultaneously have in his mind another maximal accessible variable that is related to $\theta$, but not related to $\eta$.}
\bigskip

It is crucial here that time is fixed. By letting time vary, $O$ is able to think of many variables, also unrelated ones.

In Helland (2022b, 2023b) this conclusion is applied to the observer Charlie, who observes the results of Alice and Bob in the famous Bell experiment. It is concluded that from this statement it is possible to understand that, in practice, noting that Charlie can be any observer, the violation of the CHSH inequality can be understood. Note that my conclusion here is not directly a consequence of quantum mechanics, but of a series of mathematical theorems, building upon the above four postulates.

Another application is to decision theory. Let $O$ be faced with deciding among a finite number of actions $a_1, ... ,a_r$. Define the decision variable $\theta$ as equal to $j$ if the action $a_j$ is chosen ($j=1, ... ,r$). Say that $\theta$ is accessible, and that the decision is accessible, if the decision can be carried out by $O$. The variable $\theta$ is called maximally accessible if the decision can just be carried out. Note that $\theta$ is finite-valued, so the theory of this article applies.
\bigskip

\textbf{Corollary 2}

\textit{Assume that $O$ at some fixed time $t$ has in his mind two related maximal decisions. Then he is not able to, at the same time think of another decision, which is related to the first of the two decisions, but not related to the second one.}
\bigskip

Note that here, $O$ can be any person. According to my theory, we all have this limitation in our minds. The conclusion can also be generalised to the decisions made by a communicating group of individuals.

This observation can also be used, together with the other results of this article, to give a new foundation for quantum decision theory, which I plan to discuss elsewhere.

\section{Some final remarks}

As discussed in Helland (2024a,c), other possible foundations of quantum theory have been proposed, and my approach should be compared to these. I will claim that the postulates stated here, and also in Helland (2024b,c), are simpler than most proposals in the literature, but detailed arguments behind such a claim are beyond the scope of this article, which has concentrated on the rather simple mathematics behind my approach.

A related approach, based on much more mathematics, is presented in Dutailly (2018). That article also begins with variables and derives the Hilbert space formulation from them. I argue that the topological assumptions made by Dutailly are not strictly necessary.

A limitation of my approach is that I do not assume the full validity of the superposition principle. I limit the concept of state vectors to Hilbert space vectors which are eigenvectors of some physically meaningful operators. These can be identified by questions of the form `What is $\theta$? /What will $\theta$ be if we measure it?' for some accessible variable $\theta$, together with sharp answers of the form $\theta=u$. For some such questions, answers of the type `We don't know' are allowed. Thinking in this way, gives simple explanations for paradoxes like Schr\"{o}dinger's cat and Wigner's friend; see Helland (2023c).

The postulates of this article generalise and at the same time simplify the postulates of Helland (2024c), where the symmetry conditions and the question of when a symmetric operator was self-adjoint were not taken into account. On the other hand, in Helland (2024c), conditions for the validity of the Born formula were discussed. A more thorough discussion of the Born formula is given in Helland (2024b).

\section*{References}

$\ \ \ \ \ $ Barrett, J. (2007) Information processing in generalized probabilistic theories. \textit{Physical Review} A \textbf{75}, 032304.

Busch, P. (2003). Quantum states and generalized observables: A simple proof of Gleason's theorem. \textit{Physical Review Letters} \textbf{97} (12), 120403.

Coecke, B. \& Paquette, E.O. (2009). Categories for the practicing physicist. arXiv: 0905.3010 [quant-ph].

Dutailly, J.C. (2018). Quantum mechanics revisited. arXiv: 1301.08885v3 [quant-ph].

Döring, A. and Isham, C. (2008). “What is a thing?” Topos theory in the foundations of
physics. arXiv: 0803.0417 [quant-ph].

Fuchs, C.A. and Schack, R. (2014). QBism and the Greeks: why the quantum state does not represent an element of reality. \textit{Physical Science} \textbf{90}, 015104.

Fuchs, C.A., Mermin, N.D. \& Schack, R. (2013). An introduction to QBism with an application to the locality of quantum mechanics. arXiv: 1311.5253 [quant-ph].

Hall, B.C. (2013). \textit{Quantum Theory for Mathematicians.} Springer, New York.

Helland, I.S. (2019). An epistemic interpretation and foundation of quantum theory. arXiv: 1905.06592 [quant-ph].

 Helland, I.S. (2021). \textit{Epistemic Processes. A Basis for Statistics and Quantum Theory.} Springer, Cham, Switzerland.

 Helland, I.S. (2022a). On reconstructing quantum theory from two related maximal conceptual variables. \textit{International Journal of Theoretical Physics} \textbf{61}, 69. Correction (2023) \textbf{62}, 51.

 Helland, I.S. (2022b). The Bell experiment and the limitations of actors. \textit{Foundations of Physics} \textbf{52}, 55.
 
 Helland; I.S. (2022d). On religious faith, Christianity, and the foundation of quantum mechanics. \textit{European Journal of Theology and Philosophy} \textbf{2} (1), 10-17.

 Helland, I.S. (2023a). \textit{On the Foundation of Quantum Theory. The Relevant Articles.} Eliva Press, Chisinau, Moldova.

 Helland, I.S. (2023b). An explanation of the Bell experiment. \textit{Journal of Modern and Applied Physics} \textbf{6} (2), 1-5.

 Helland, I.S. (2023c). Possible connections between relativity theory and a version of quantum theory based upon theoretical variables. arXiv: 2305.15435 [physics.hist-ph].
  
 Helland, I.S. (2023d). Quantum mechanics as a theory that is consistent with the existence of God. \textit{Dialogo Conferences and Journal} \textbf{10} (1), 127.134.
 
 Helland, I.S. (2023e). A simple quantum model linked to decisions. \textit{Foundations of Physics} \textbf{53}, 12.
  
 Helland, I.S. (2024a). An alternative foundation of quantum mechanics. arXiv: 2305.06727 [quant-ph]. \textit{Foundations of Physics} \textbf{54}, 3.
 
 Helland, I.S. (2024b). On probabilities in quantum mechanics. arXiv: 2401.17717 [quant-ph]. \textit{APL Quantum} \textbf{1}, 036116.
 
 Helland, I.S. (2024c). A new approach towards quantum foundation and some consequences. arXiv: 2403.09224 [quant-ph]. Academia Quantum 1. https://doi.org/10.20935/AcadQuant7282.
 
 Helland, I.S. (2025). Quantum probability for statisticians; some new ideas. arXiv: 2503.02658 [quant-ph]. Submitted.

 Helland, I.S. \& Parthasarathy, H. (2024). \textit{Theoretical Variable, Quantum Theory, Relativistic Quantum Field Theory, and Quantum Gravity.} Manakin Press, New Dehli.
 
 Jodlicka, P.O., Kos, S., Smid, Vomlel, J., and Slavik (2025). Has anything changed? Tracking long-term interpretational preference in quantum mechanics. arXiv: 2507.09988 [quant-ph].
 
 Khrennikov, A. (2010). \textit{Ubiquitus Quantum Structure. From Psychology to Finance.} Springer, New York.
 
 Khrennikov, A. (2023). \textit{Open Quantum Systems in Biology, Cognitive and Social Sciences.} Springer, Cham, Switzerland.
 
 Khrennikov, A. (2024a). Ozawa's intersubjectivity theorem as objection to QBism individual agent perspective. \textit{International Journal of Theoretical Physics} \textbf{63}, 23.
 
 Khrennikov, A. (2024b).Relational quantum mechanics: Ozawa's intersubjectivity theorem as justification of the postulate on internally consistent descriptions. \textit{Foundations of Physics} \textbf{54}, 29.
 
 Khrennikov, A., Ozawa, M., Benninger, F., and Schor, O. (2025). Coupling quantum-like cognition with neuronal netwoorks within generalized probability theory. \textit{Journal of Mathematical Psychology} \textbf{125}, 102923.
 
 Klein, U. (2010). The statistical origins of quantum mechanics. \textit{Hindawi Publishing Corperation Physics Physics Research International} \textbf{2010}, 808424.
 
  M\"{u}ller, M.P. (2021). Probabilistic theories and reconstructions of quantum theory. \textit{SciPost Physical Lecture Notes} \textbf{28}, 1-41.
  
  Ozawa, M. (2024). Intersubjectivity and value reproducibility of outcomes of quantum measurements. arXiv: 1911.10893 [physics, gen-ph].

 .

\end{document}